\documentclass[twocolumn]{aastex61}
\pdfoutput=1 
\usepackage{amsmath,amstext}
\usepackage[T1]{fontenc}
\usepackage{apjfonts} 
\usepackage[figure,figure*]{hypcap}
\usepackage{gensymb}

\shorttitle{Observations of the Kelvin-Helmholtz instability}
\shortauthors{Hiller et al.}

\begin{document}

\title{Observations of the Kelvin-Helmholtz instability driven by dynamic motions in a solar prominence}

\author{Andrew Hillier}
\affiliation{College of Engineering, Mathematics and Physical Sciences, University of Exeter, Exeter, EX4 4QF UK}
\author{Vanessa Polito}
\affiliation{Harvard-Smithsonian Center for Astrophysics, 60 Garden Street, Cambridge, MA 02138, USA.}

\begin{abstract}
Prominences are incredibly dynamic across the whole range of their observable spatial scales, with observations revealing gravity-driven fluid instabilities, waves, and turbulence. With all these complex motions, it would be expected that instabilities driven by shear in the internal fluid motions would develop. However, evidence of these have been lacking. Here we present the discovery in a prominence, using observations from the Interface Region Imaging Spectrograph (IRIS), of a shear flow instability, {the Kelvin-Helmholtz sinusoidal-mode of a fluid channel}, driven by {flows} in the prominence body. This finding presents a new mechanism through which we can create turbulent motions from the flows observed in quiescent prominences. The observation of this instability in a prominence highlights their great value as a laboratory for understanding the complex interplay between magnetic fields and fluid flows that play a crucial role in a vast range of astrophysical systems.
\end{abstract}


\section{Introduction}

Solar prominences are cool plasma suspended in the $10^6$\,K solar corona by magnetic fields \citep{TAN1995}. Space-based observations has revolutionised our understanding of prominences, where we now know that they are incredibly dynamic across the whole range of observable spatial scales \citep{MAC2010}. Investigations show that the dynamical features observed in prominences both drive and are driven by gravity-driven fluid instabilities \citep{BERG2010, BERG2011, HILL2011b, HILL2012, HILL2018}, waves \citep{ARRE2012, HILL2013, ANT2015}, and turbulence \citep{LEO2012, FREED2016, HILL2017}. 

The complex motions observed in prominences can be clearly seen to create shear flows, and so it would be expected that instabilities driven by this shear would develop. The classic shear flow instability is the Kelvin-Helmholtz instability (KHi), which breaks up coherent sheets of vorticity into vortices. This instability comes in two distinct flavours: the surface mode of the instability that drives vortex formation at the boundary between two non-parallel flows\citep{CHAN1961}, and {the modes that act on channel flows including the sinusoidal-mode which drives the development of serpentine patterns} \citep{DRAZINREID1981}. {Magnetic fields work to suppress the instability. For an arbitrary shear flow, stability of the flow is guaranteed unless the difference between the maximum and minimum velocities is twice the minimum Alfv\'{e}n velocity in the direction of the flow \citep{HUGHES2001}.}

{The surface mode of the KHi has been observed in many astrophysical systems. This includes where the solar wind interacts with the flanks of the magnetosphere \citep[e.g.][]{HASE2004}, associated with erupting regions \citep{OFMAN2011}, on the flanks of coronal mass ejections \citep{FOU2011, MOSTL2013} and where emerging magnetic flux interacts with prominences \citep[e.g.][]{BERG2010, RYU2010, BERG2017}. The sinusoidal-mode of a channel flow has proved more elusive, but it is believed to be important in coronal plumes \citep{ANDR2001}, and astrophysical jets \citep{FERR1981}.}

{There has been a wide range of numerical and analytical studies investigating the role of this instability in astrophysical settings, often in the context of explaining observations \citep[e.g][]{FOU2011, OFMAN2011, MOSTL2013}. 
\citet{MIURA1982} investigated the linear growth rate of the magnetic KHi for continuous compressible flows finding that increases in the width of the shear layer reduces the growth of the instability and that the instability can be suppressed by compressibility. 
One important role of the nonlinear evolution of the KHi is its ability to develop turbulent flows through reconnection and secondary instabilities \citep[e.g.][]{MATSUMOTO2004}.
This process has been seen in numerical studies of kink waves in the solar atmosphere, which found that the KHi can grow and become turbulent on the surface of oscillating coronal loops \citep[e.g.][]{TERR2008} and prominence threads \citep[e.g.][]{ANT2015}.}

{There have been observations in the solar atmosphere, including in prominences, of the surface mode of the magnetic KHi 
 but observations of the {sinusoidal-mode} of the instability are still lacking. 
Here we present the discovery in a prominence, using observations from the Interface Region Imaging Spectrograph \citep[IRIS;][]{DEPON2014}, of streams of fluid developing serpentine patterns as a result of becoming unstable to the KHi. }

\section{Observations}

\begin{figure*}
\begin{center}
\includegraphics[width=0.95\textwidth]{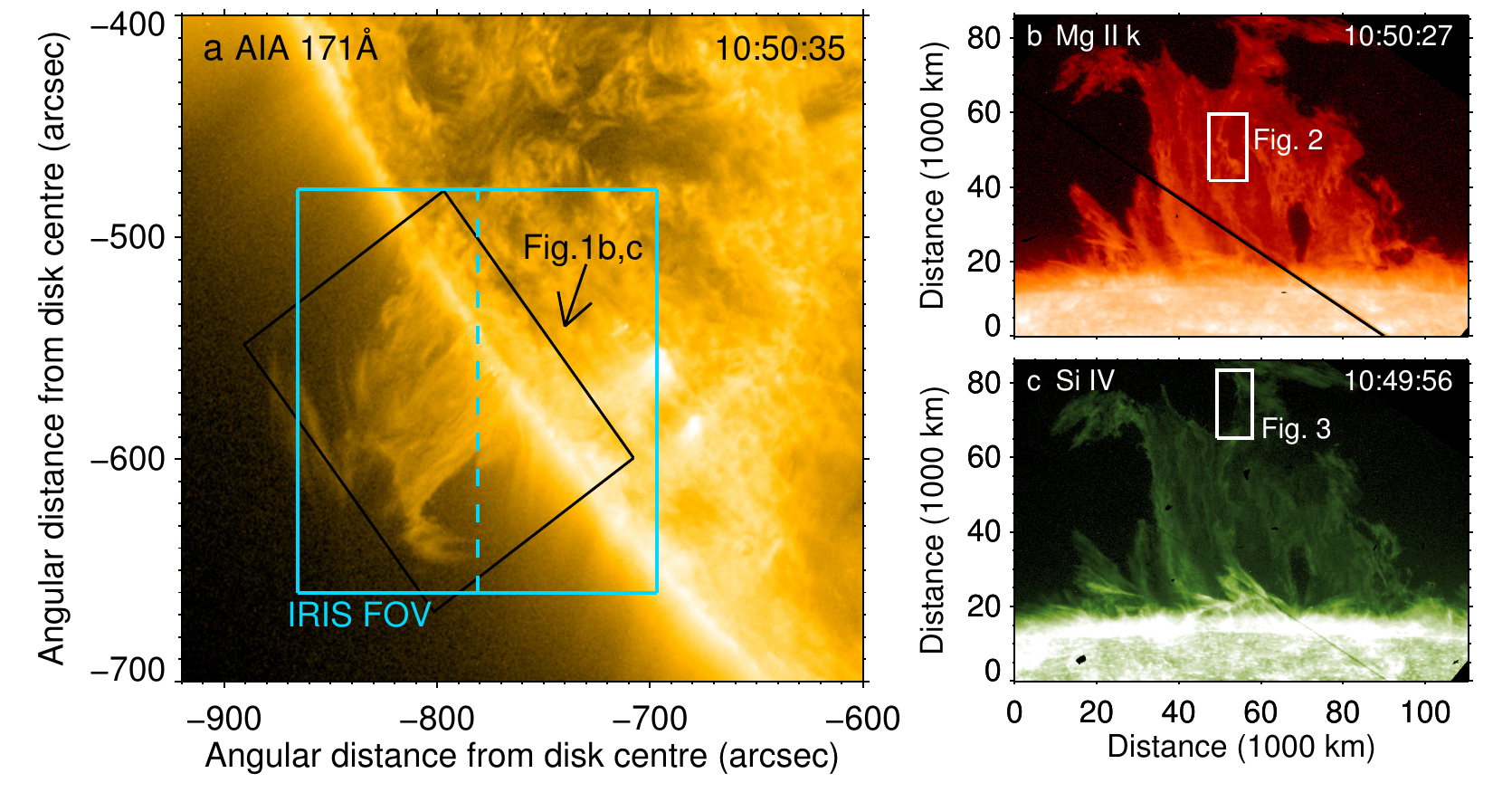}
\caption{Quiescent prominence observed by IRIS and AIA on 30 June 2015. Panel {\bf a} shows the 171 {\AA} image with the intensity in log-scale.
 The light blue box shows the field of view of IRIS and the slit location (dotted line). The rotated black box indicates the prominence region shown in panels {\bf b} and {\bf c}. 
Panels {\bf b} and {\bf c} show the \ion{Mg}{2} k and \ion{Si}{4} images of the prominence (log-scale) with the white boxes marking the regions used in Fig.\,\ref{fig2} and \ref{fig3} respectively.
The black line in panels b-c shows the position of the slit. Movies available from the published paper.}
\end{center}
\label{fig1}
\end{figure*}

On 30 June 2015, IRIS observed a quiescent prominence on the Southeast limb between 6:57 and 11:20 UT (Fig.\,\ref{fig1}). The IRIS slit-jaw imager (SJI) observed the prominence in three broadband filters centred on the \ion{Mg}{2} k 2796.4\,{\AA}, \ion{C}{2} 1335.78\,{\AA} and \ion{Si}{4} 1402.77\,{\AA} lines, formed at temperatures of $10^4$, $10^{4.5}$ and $10^{4.9}$\,K respectively, making it perfect for the study of prominence dynamics \citep[e.g.][]{SCH2014}. The images were taken with a $96$\,s cadence over a field-of-view of $167$''$\times174$'' and a spatial resolution of approximately 120 km (0.167''/pix). 
The IRIS spectrograph ran a sit-and-stare study, {whose slit position is} marked by the black line in Fig.\,\ref{fig1}).
While the IRIS slit was observing a relatively quiet region of the prominence, the IRIS SJI images caught interesting dynamics in the prominence body. 
{There is no Hinode SOT data available for this study.}

The Solar Dynamics Observatory (SDO) Atmospheric Imaging Assembly \citep[AIA;][]{LEM2012} acquired full-Sun images with a spatial resolution of approximately $870$--$1,230$\,km (0.6''/pix) and a cadence of $\sim12$\,s.
In this work, we use 171\,{\AA} {images}, mostly showing $0.7$--$1.5\times10^6$\,K optically thin emission, to provide contextual information on the coronal plasma in and around the prominence (Fig.\,\ref{fig1}). 

To highlight the dynamical features under study an unsharpened mask was applied to the IRIS SJI images, as shown in Figs.\,\ref{fig2} \& \ref{fig3}. 
We co-aligned the IRIS and AIA observations to correct for small differences in the instruments pointing. 
The co-alignment was performed by {comparing} images of the prominence formed in the IRIS SJI \ion{Mg}{2} and \ion{Si}{4} passbands with the AIA 304 {\AA} images (T$=10^{4.7}$\,K).

The prominence reached a height of approximately 55,000 km above the solar limb with a width of 60,000 km and was relatively square in shape (see Fig.\,\ref{fig1}). What appears to be a bubble \citep{BERG2010} was visible in the lower right region of the prominence and the main body of the prominence consisted of multiple, recurring flows both aligned with \citep[e.g.][]{CHAE2010} and in the opposite direction \citep[e.g.][]{HILL2011a} to solar gravity. The central region of the prominence presented many clear examples of these flows. Close investigation of the flows in this region reveal that they develop a shear-flow instability and roll up on themselves through the formation of vortices (see Figs.\,\ref{fig2} and \ref{fig3}), although, due to the placement of these flows far from the IRIS slit position, no spectral data is available. We present two of the clearest examples of these dynamics. 

\section{Analysis}\label{analysis}

\begin{figure}
\begin{center}
\includegraphics[width=0.435\textwidth]{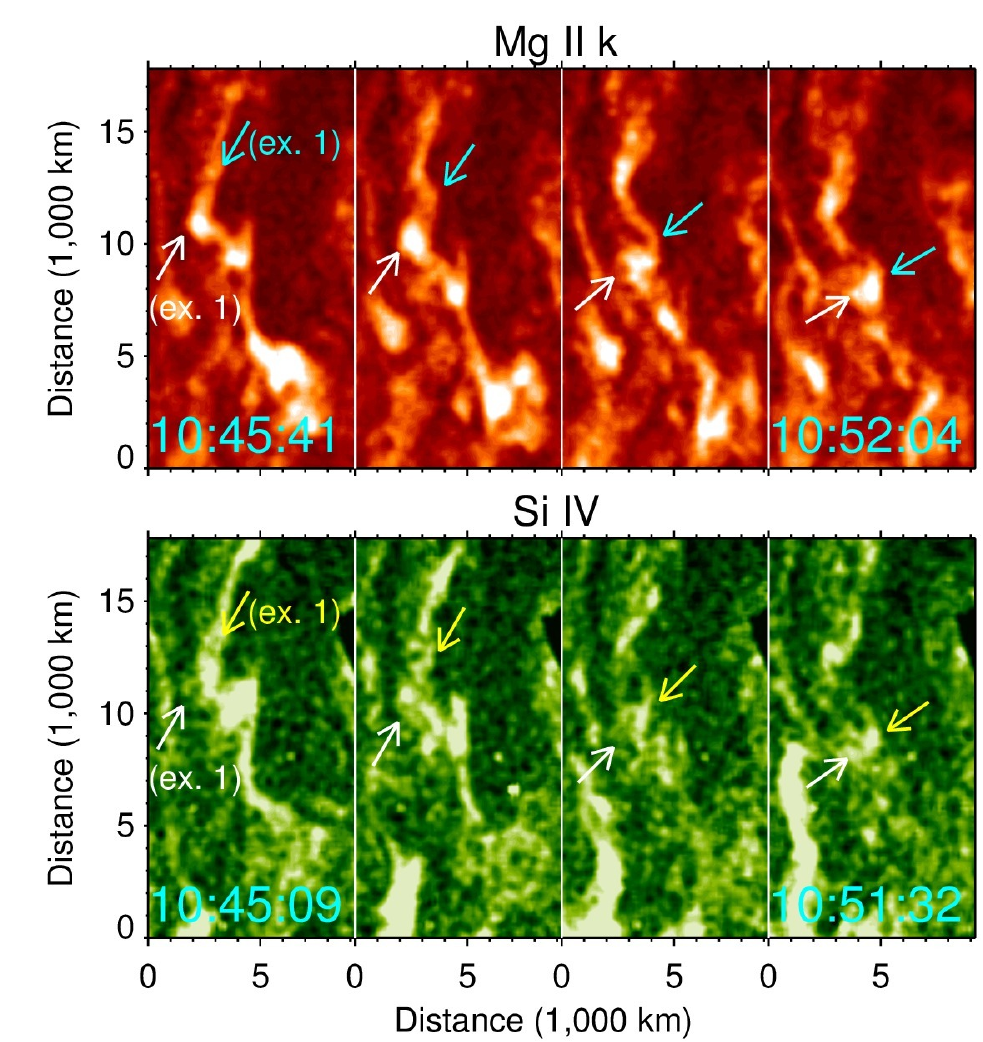}\\
\includegraphics[width=0.435\textwidth]{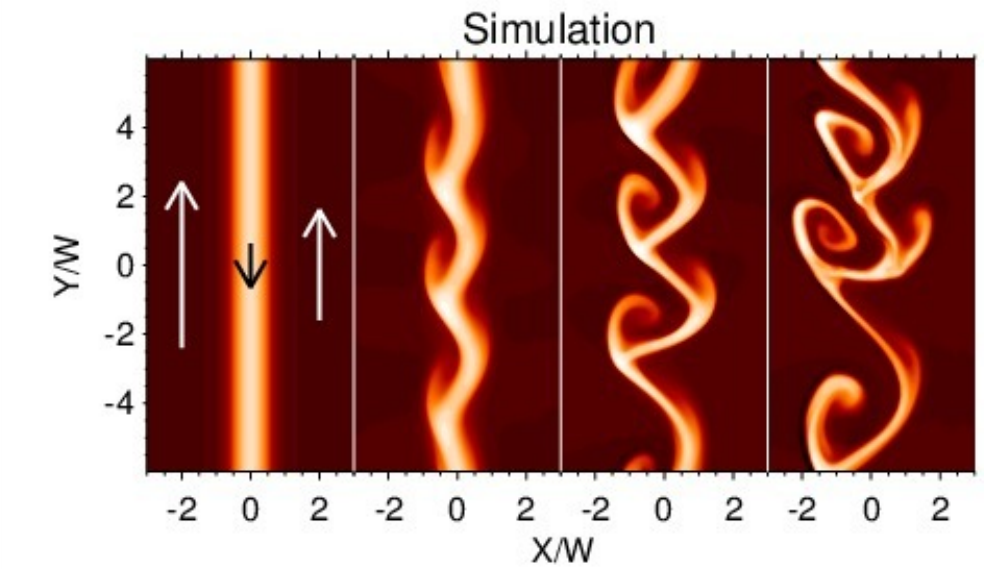}
\caption{Top and middle rows: series of images showing the rolling up of a downflow in the \ion{Mg}{2} k and \ion{Si}{4} channels respectively. The observed intensities are taken from the region marked in Fig\,\ref{fig1}, panel b. The arrows marked ex.\,1 show the example of the downflow that rolls up and are used to highlight the rotation of the structure. The start and end time in UT of is given at the base of the images. \\Bottom rows: MHD simulation for ex. 1 using a cadence of 320\,s. The arrows in the first panel show the initial velocity distribution. $W$ is the flow width.}
\end{center}
\label{fig2}
\end{figure}

One of the downflows, observed in the IRIS SJI \ion{Mg}{2} k channel, forms our example 1 (Fig.\,\ref{fig2}, top two rows). This particular flow accelerated at approximately one quarter of solar gravitational acceleration, from 9\,km/s to a speed of 16\,km/s. It had a width of 900 km and is characterised by a bright, descending plasma blob as part of a chain of blobs moving downwards in a bright thread. As this blob falls, its tail bellows out to the right (as viewed by the observer) and begins to wrap around the structure, the whole process taking around 360\,s. The length along the thread associated with the rolling up of the downflow is 3,200\,km.

\begin{figure}
\begin{center}
\includegraphics[width=0.435\textwidth]{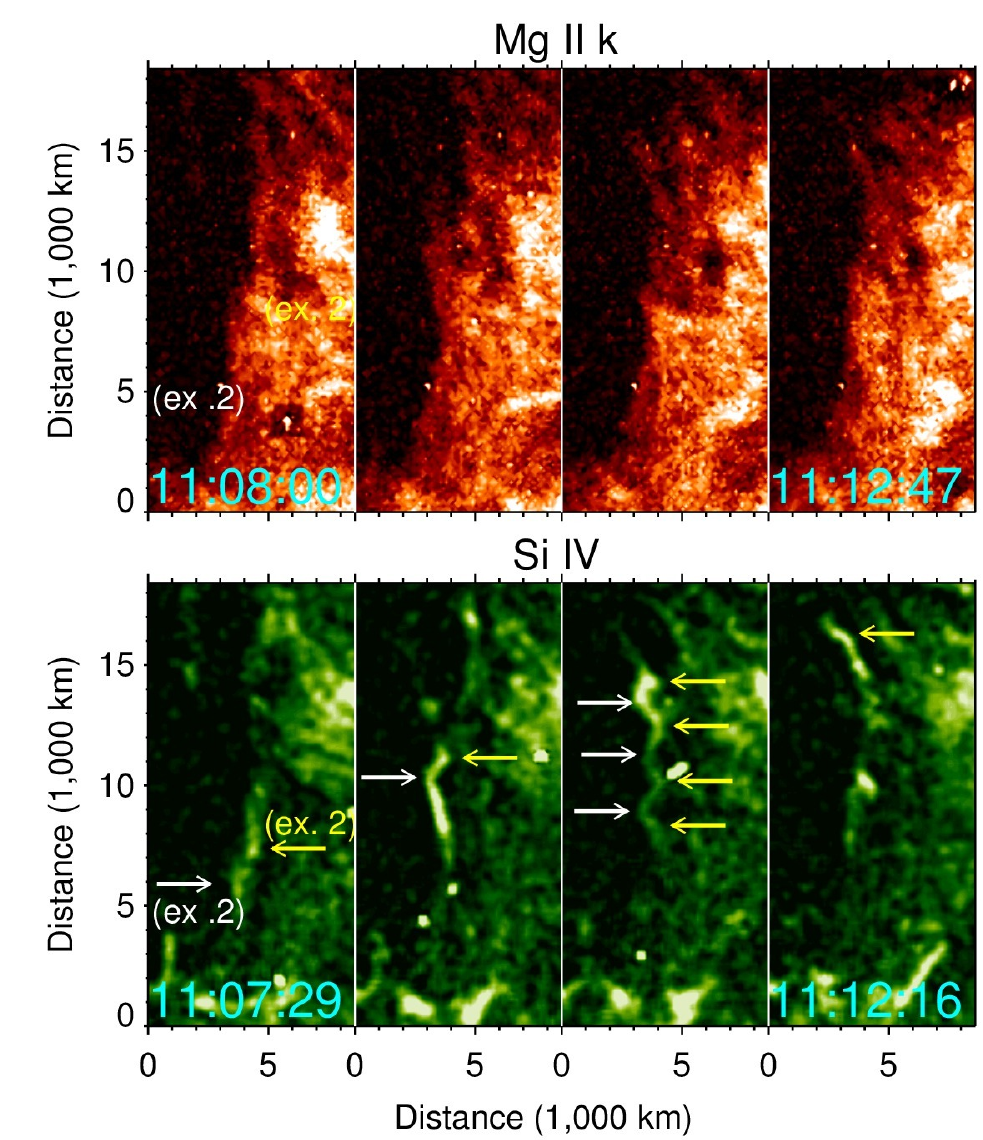}\\
\includegraphics[width=0.435\textwidth]{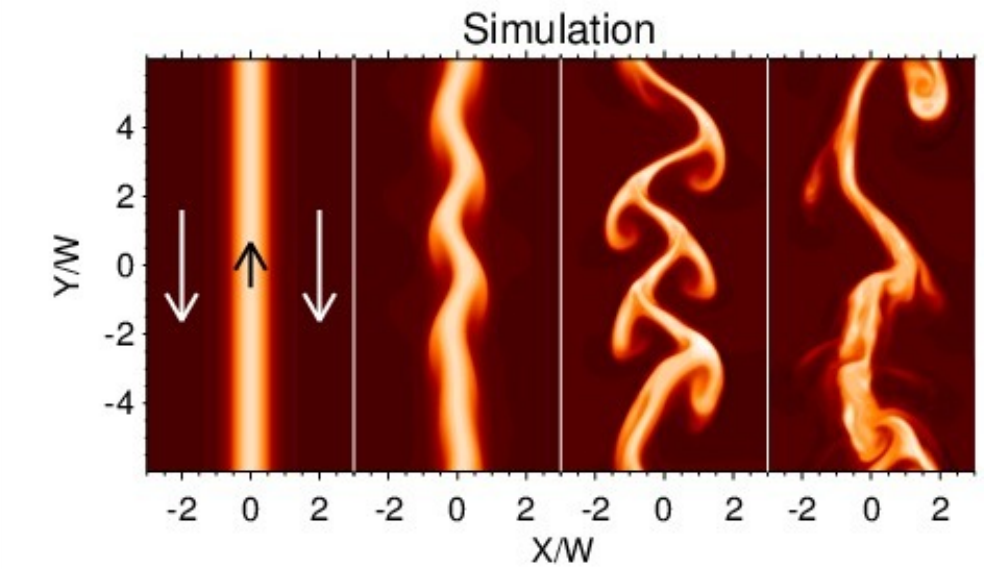}
\caption{Top and middle rows: series of images showing an unstable upflow in the \ion{Mg}{2} k and \ion{Si}{4} channels respectively. The observed intensities are taken from the region marked in Fig\,\ref{fig1}, panel c. By the third panel, three undulations have developed. The start and end time in UT of is given at the base of the images. \\Bottom rows: MHD simulation for ex. 2 using a cadence of 204\,s. The arrows in the first panel show the initial velocity distribution. $W$ is the flow width.}
\end{center}
\label{fig3}
\end{figure}

The second downflow (Fig.\,\ref{fig3}, top two rows) was found in a warm ($\approx 8\times10^4$\,K) ejection that formed part of a stream of upward moving plasma observed in the IRIS SJI Si IV channel. The ejected thread, with width of 480\,km, moved upward at a projected speed of 34\,km/s. It did not remain straight, but develops a sinusoidal pattern that is symmetric about the axis of the thread. This develops incredibly quickly, i.e. over the space of 90\,s, and disappears just as rapidly. The wavelength of sinusoidal structure is 2,000\,km. 

There are two interesting differences between these two examples. Firstly, the second unstable serpentine structure is relatively symmetric about its central axis, which is not the case for the first example. This suggests that different flow patterns are possibly at play in the two cases. In addition, while the first downflow is noticeably clearer in the chromospheric \ion{Mg}{2} k line, the second downflow is only observed in the TR \ion{Si}{4} passband, highlighting the different temperatures at which these dynamics occur. Not only does this tell us that observing prominences across a wide range of temperatures is important to reveal the full range of motions, but also that there are not any clear temperature changes during the evolution of the flow.

\begin{figure*}
\begin{center}
\includegraphics[width=0.435\textwidth]{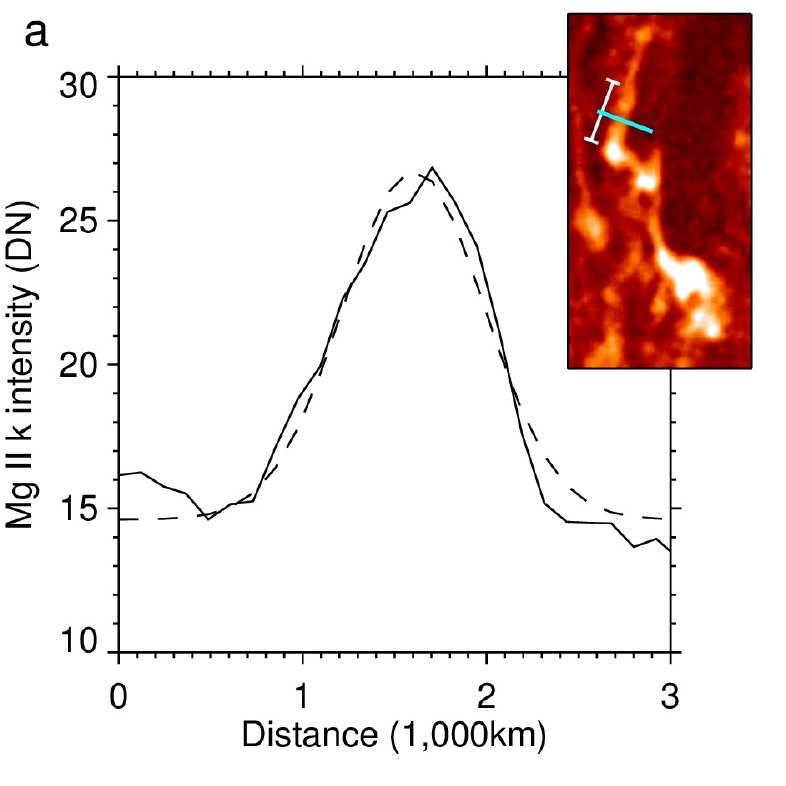}
\includegraphics[width=0.435\textwidth]{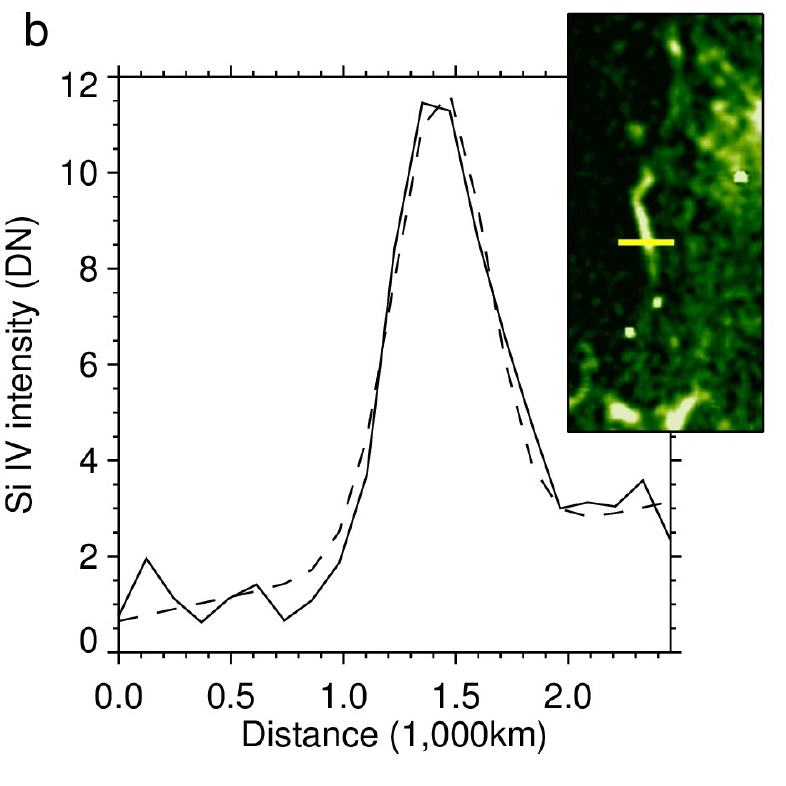}
\caption{Distribution of intensity across the threads for ex. 1 ({\bf a}) and ex. 2 ({\bf b}). The intensity (solid line) is  fitted with a Gaussian distribution (dashed line). The inset in each panel shows the position of the slit used to measure the intensity across the thread (turquoise for {\bf a} and yellow for {\bf b}). The white line in the panel {\bf a} inset gives the wavelength.}
\end{center}
\label{fig4}
\end{figure*}

The main quantities calculated from the data in this paper are the speeds of the flows, their widths and the wavelengths of the instabilities. The speed was calculated by determining the change of position of the bright structure in the thread between images and dividing this distance by the cadence ($96$\,s). The method for determining the thickness of the thread, shown in Fig.\,\ref{fig4}, is based on the full-width-half-maximum of a fitted Gaussian distribution to the intensity across the thread and taking this as the thread thickness. For example 1 the estimate of the wavelength is shown in Fig.\,\ref{fig4} panel a, whereas for example 2 it is calculated as the mean distance between the wavelength peaks in Fig.\,\ref{fig3}.

\section{Simulations}

To attempt to model the observed dynamics we performed numerical simulations using the (P{\underline I}P) code \citep{HILL2016}.
The simulation presented here is a 2.5D calculation of a plasma $\beta=0.3$ and $\gamma=5/3$ medium with the magnetic field at an angle of $5^\circ$ to the normal of the plane resulting in $B_y$ and $B_z$ components{, which is consistent with measurements of the predominantly horizontal magnetic field, low plasma $\beta$ environment of quiescent prominences \citep[see][and references therein]{MAC2010}.}
The simulation domain is $-4$ to $4$ in the x direction and $-6$ to $6$ in the y direction {(where these lengths are normalised by the width of the jet ($W$) and allow at least two wavelengths of the instability to form)} with resolution of $200\times300$ grid points. 

There are two different initial conditions used for the numerical experiments.
The initial density and velocity profiles are given as:
\begin{align}
\rho=\rho_{\rm l}+&\frac{1}{4}(\rho_{\rm u}-\rho_{\rm l})\left(\tanh\left[\left(\frac{0.5-x}{0.3}\right)\right]+1.0\right)\\ &\times\left(\tanh\left[\frac{x+0.5}{0.3}\right]+1.0\right)\nonumber\\
v_y=v_{y\rm l}+&\frac{1}{A}(v_{y\rm u}-v_{y\rm l})\left(\tanh\left[\left(\frac{0.5-x}{0.3}\right)\right]+B\right)\label{sim_flow}\\ &\times\left(\tanh\left[\frac{x+0.5}{0.3}\right]+1.0\right)\nonumber
\end{align}
where $\rho_{\rm l}=1$, $\rho_{\rm u}=2$, $v_{y\rm l}=-0.8C_{\rm S}$ and $v_{y\rm u}=0.4C_{\rm S}$, with $A=6$ and $B=2$ for example 1 and $A=4$ and $B=1$ for example 2.
Equation \ref{sim_flow} produces the velocity profile depicted by the arrows in the first panel of the bottom rows of Figs. \ref{fig2} and \ref{fig3}.
Note that the values of the velocity are chosen to make the linear instability development static in the rest frame of the simulation, and to match with the observations we use a shear flow velocity that is close to the value of the local sound speed (e.g. approximately 10 km/s in the dense, cool regions of the prominence). {The density variation is taken to be within the expected variation of prominence density \citep[e.g.][]{LAB2010}}.
The instability is initiated with a small amplitude random noise perturbation in the velocity field.
{Gravity is not included in these calculations, but as the vertical direction is orthogonal to the density gradients its inclusion would not change the onset of instability. } 

During the evolution of the density distribution for both simulations (see bottom rows of Figs. \ref{fig2} and \ref{fig3}), the dense thread becomes unstable and forms undulations.
This is as a result of the formation of alternating vortices on either side of the thread and leads to structures that are visually similar to those observed. Once the instability is sufficiently evolved, currents build up as a consequence of the bending of the magnetic field, which ultimately results in the magnetic field reconnecting and destroying the undulations \citep{MAK2016}. This physical process reproduces the key features of the observed dynamics. The main difference between the two simulations is that the first has the largest velocity on the left hand side, while the velocities on the left and right are the same in the second simulation. While the former simulation leads to the billowing out to the right as observed in Fig.\,\ref{fig2}, the symmetric flow leads to symmetric undulations as seen in Fig.\,\ref{fig3}.
{In the non-linear stage of the simulations there are features that we cannot readily identify in the observations, including the spurs that extrude from the sinusoidal shapes in Fig.\,\ref{fig3}. The nonlinear dynamics will depend on the parameters of the system and we expect stronger magnetic fields, for example, would reduce these spurs. A full parameter study would reveal the best parameters to reproduce the observations, though higher spatial and temporal resolution of the observations may reveal these structures occurring in the prominence.}

The simulated instability, which so nicely matches with the observed dynamics, is a modified version of 
 the Bickley jet. 
{This} jet is unstable to the {sinusoidal-mode} of the KHi \citep{DRAZINREID1981}, which is driving the undulations in our simulations {by making} the jet develop a sinusoidal structure at a preferred wavelength of $\sim3.5$ times the characteristic width of the flow \citep{DRAZINREID1981, HUGHES2001}. For the observations, we find aspect ratios of the width and the wavelength of 3.5 for the example 1 and 4.1 for example 2, which are similar to the value predicted by theory.

{An analytical statement of the linear stability criterion of our model in not possible, but it is for the simplified setting of an incompressible slab (mimicking the dense thread) symmetric about $x=0$ of width $W$ with discontinuous jumps in the density and velocity and a uniform magnetic field. 
For $2$D perturbations to this model the dispersion relation is given by \citep[e.g.][]{NAKA1996}:
\begin{align}
\rho_2\left((V_{y,2}+c)^2-V_{{\rm A},y,2}^2\right)&\tanh\left(\frac{kW}{2}\right) \\+&\rho_1\left((V_{y,1}+c)^2-V_{{\rm A},y,1}^2\right)=0\nonumber
\end{align}
where the subscripts $1$ and $2$ denote the external and internal medium respectively for the velocity ($V_y$), density ($\rho$) and Alfv\'{e}n speed in the {vertical} direction ($V_{{\rm A},y}$), and $c$ is the complex wave speed.
This gives a condition for instability, assuming $\rho_2>\rho_1$, of:
\begin{equation}
M_{\rm A}>\sqrt{\frac{\rho_1}{\rho_2}}+1,
\end{equation}
where $M_{\rm A}=|V_{y,2}-V_{y,1}|/V_{{\rm A},y,1}$. For small density contrasts this becomes $M_{\rm A}\gtrsim 2$.
However, it is not clear by how much the instability bounds would increase for the non-discontinuous profiles expected in the solar atmosphere.}

{We have focussed on the interpretation that the KHi drives the dynamics. However a couple of related explanations may exist. An MHD kink wave would produce sinusoidal structures as would a negative energy wave \citep[a dissipative instability giving an overstable kink wave, e.g.][]{RUDE1996}. Example 1 shows direct instability growth so these other explanations do not hold, but for example 2 the low cadence of the observations means we have to consider these possibilities. If this were a wave, it'd have to be highly nonlinear though there is no obvious strong driver (a negative energy wave may circumnavigate this) and there are no observed oscillations. Also a wave would struggle to explain why these structures only develop for shear flows and with the particular aspect ratio. Higher cadence observations would help to properly discount these possibilities.}

\section{Discussion}

One question that needs to be addressed is: in highly dynamic prominences, why do we not see this instability developing everywhere? In some cases it will just be that the angles of the instability to the line of sight are such that nothing can be observed even though the instability is growing. However, one part of this answer is likely to be that the magnetic fields are strong enough to at least delay the growth of the instability so that it does not become noticeable in many places. 
It is quite likely that, now we have discovered the presence of the KHi associated with prominence flows, many more instances will come to light. 

{In these observations we found a shear flow instability for a velocity differences {between $\sim 15$ and $35$\,km\,s$^{-1}$} {and for this to be unstable we require $M_{\rm A}\gtrsim 2$.}
As this condition does not include other important suppression mechanisms like continuous distributions, compression and viscosity (though the large Reynolds number will make this relatively unimportant), we can expect that the actual requirement for instability will be greater.
Therefore, to achieve the fast development the instability as presented here velocity differences somewhere beyond this limit will be required.}

An interesting result is that 
the flows presented here cannot be moving along magnetic field lines, unless the magnetic field strengths of this prominence are weaker than measured values \citep[e.g.][]{LEROY1989, CASINI2009, OROZCO2014, LEVENS2016}, but are moving almost perpendicular to the field, carrying it along with the flow. 
The high frequency with which small-scale flows appear means that the prominence magnetic field must be constantly twisted up and redistributed by these flows. Over the lifetime of a prominence this may lead to magnetic energy being transferred around the prominence, slowly evolving the global structure until it becomes unstable and erupts.

{In our modelling we have not taken into account the optically thick prominence \ion{Mg}{2} emission.
Though the characteristic speeds used in the modelling in this paper can be easily obtained by tracking prominence motions and the size of the flows can be measured, it is much harder to accurately determine the velocity and density distributions. 
Therefore, more work is necessary to go beyond the proof-of-concept simulations presented in this paper and to make it to possible to use this instability to directly infer the plasma and magnetic field conditions inside the prominence.}

The discovery of this instability provides an explanation for how observed vortical motions \citep{LIGG1984} can be formed in a prominence and also has great implications for understanding the development of turbulence in prominences. 
Investigations into this turbulence have revealed that a characteristic length scale of a few thousands of kilometres exists in the turbulence \citep{LEO2012, HILL2017}, where this length is a factor of a few longer in the vertical direction than the horizontal \citep{HILL2017}. The unstable flows presented here are at just the right size to drive turbulence from this scale inside the body of the prominence. 

The magnetic KHi is one of the fundamental instabilities of fluid dynamics, and this means that it regularly occurs 
in many different astrophysical systems across a huge range of scales. Due to the commonality of the physics, by investigating this instability in one system we can learn about how it works in a wide range of other astrophysical systems. The high temporal and spatial resolution that is given by space-based, and will be given by future ground-based, observations of prominences means that we have an exceptional opportunity to investigate this important astrophysical phenomenon.

\acknowledgments
IRIS is a NASA small explorer mission developed and operated by LMSAL with mission operations executed at NASA Ames Research center and major contributions to downlink communications funded by ESA and the Norwegian Space Centre. 
AIA data are courtesy of NASA/SDO and the AIA science team. 
The authors would like to thank Dr. H. Mason and Dr. G. Del Zanna for insightful discussions. 
A.H. is supported by his STFC Ernest Rutherford Fellowship (ST/L00397X/2).  
V.P. is supported by NASA grant NNX15AF50G, and by contract $8100002705$ from Lockheed--Martin to SAO. \\

\end{document}